\documentclass[11pt,twoside]{article}

\usepackage{asp2006}
\usepackage{epsf}
\usepackage{psfig}
\usepackage{lscape}
\usepackage{graphicx}

\begin{document}

\title{Large-Scale Alignments of Quasar Polarization Vectors: Evidence
at Cosmological Scales for Very Light Pseudoscalar Particles Mixing
with Photons ?}

\author{D. Hutsem\'ekers (1$^{\star}$), A. Payez (1$^{\dagger}$),
R. Cabanac (2), H. Lamy (3), D. Sluse (4), B. Borguet (1), and
J.-R. Cudell (1)}

\affil{(1) University of Li\`ege, Belgium , (2) Observatoire
Midi-Pyr\'en\'ees, France, (3) BIRA-IASB, Belgium, (4) EPFL, Switzerland}

\renewcommand{\thefootnote}{\fnsymbol{footnote}}
\footnotetext[1]{Senior Research Associate FNRS, $^{\dagger}$IISN Research Fellow}
\renewcommand{\thefootnote}{\arabic{footnote}}
\setcounter{footnote}{0}

\begin{abstract}
Based on a sample of 355 quasars with significant optical
polarization, we found that quasar polarization vectors are not
randomly oriented over the sky as naturally expected. The probability
that the observed distribution of polarization angles is due to chance
is lower than 0.1\%. The polarization vectors of the light from
quasars are aligned although the sources span huge regions of the sky
($\sim$ 1~Gpc). Groups of quasars located along similar lines of sight
but at different redshifts (typically z $\sim$ 0.5 and z $\sim$ 1.5)
are characterized by different preferred directions of
polarization. These characteristics make the observed alignment effect
difficult to explain in terms of a local contamination by interstellar
polarization in our Galaxy. Interpreted in terms of a
cosmological-size effect, we show that the dichroism and birefringence
predicted by a mixing between photons and very light pseudoscalar
particles within a magnetic field can qualitatively reproduce the
observations. We find that circular polarization measurements could
help constrain this mechanism.
\end{abstract}

\section{Introduction}
Large-scale alignments of quasar polarization vectors were first
uncovered looking at a sample of 170 quasars selected from the
literature (Hutsem\'ekers 1998, hereafter Paper I). The presence of
such alignments was confirmed later on a larger sample (Hutsem\'ekers
\& Lamy 2001, hereafter Paper II). The departure from random
orientations was found at significance levels small enough to merit
further investigation. Moreover, these alignments seemed to come from
high redshift regions, implying that the underlying mechanism might
cover physical distances of gigaparsecs. A large survey of linear
polarization was then started, with the goal to characterize better
the polarization properties of quasars and to investigate the reality
of the alignments. A final sample of 355 quasars with reliable
polarization measurements was then built on the basis of the new
polarization measurements and of a comprehensive compilation from the
literature. A detailed analysis of this sample was carried out by
Hutsem\'ekers et al. (2005, hereafter Paper III). The main results are
reviewed here and a possible interpretation based on
photon-pseudoscalar mixing is presented.

\section{Large-scale coherent orientations of  quasar polarization vectors}
\begin{figure}[!p]
\resizebox{\hsize}{!}{\includegraphics*{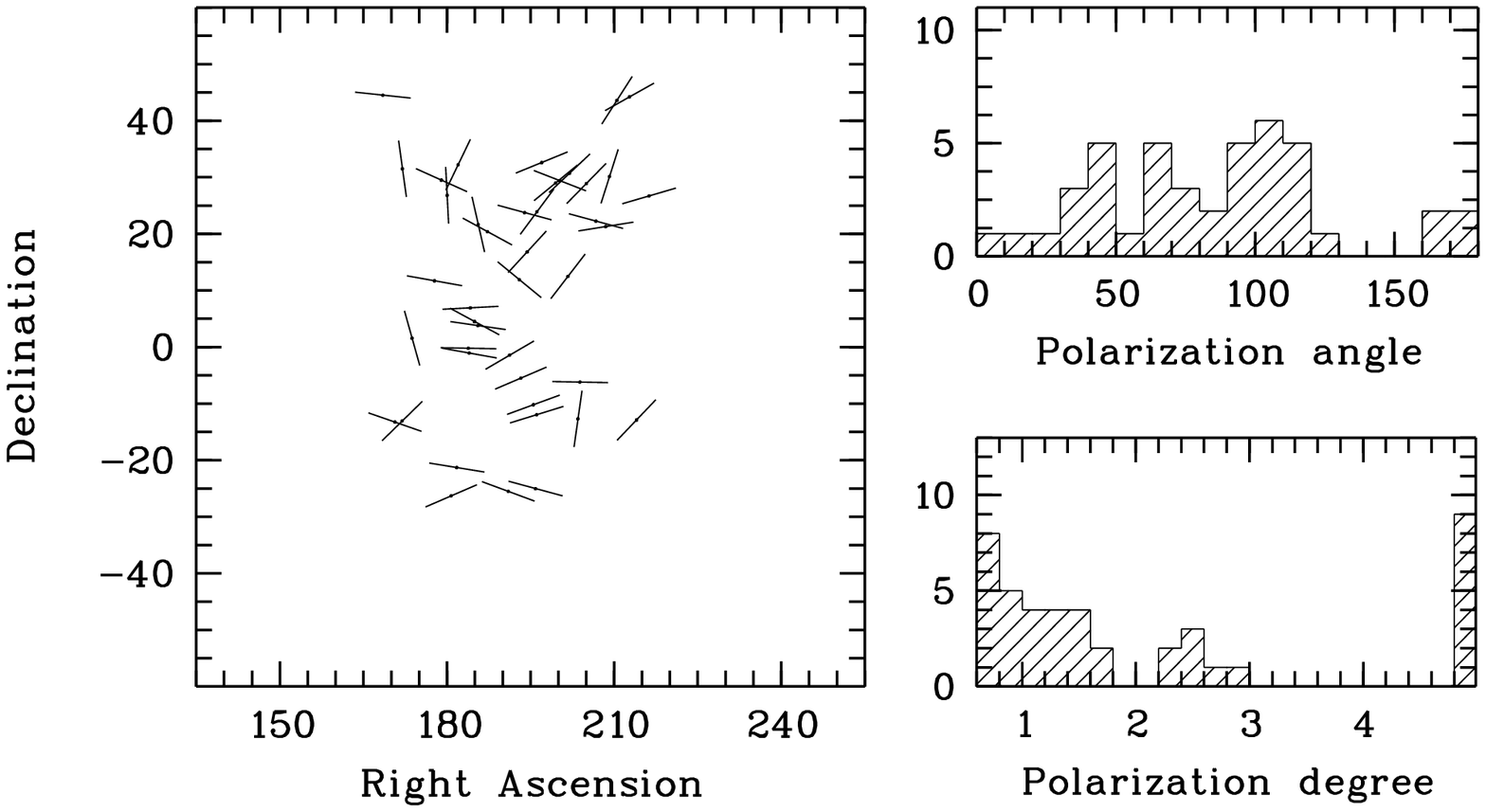}}\\[0.4cm]
\resizebox{\hsize}{!}{\includegraphics*{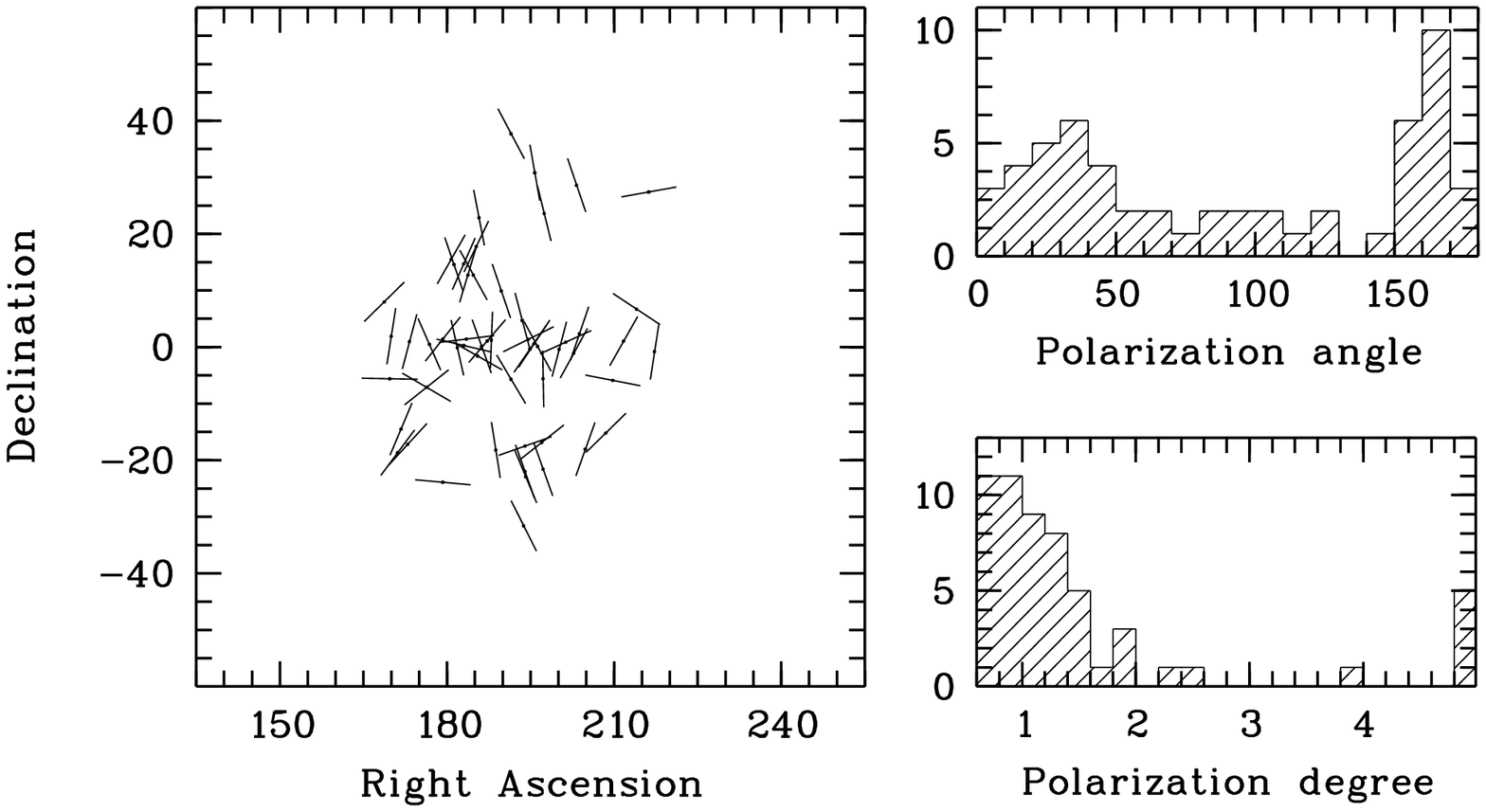}}
\caption{Maps of quasar polarization vectors towards the North
Galactic Pole region, together with the corresponding distributions of
polarization angle (in $^o$) and polarization degree (in \%). The
regions illustrated are delimited in right ascension and declination
by $168^{o} \leq \alpha \leq 218^{o}$ and $\delta \leq 50^{o}$, and in
redshift by $0.0 \leq z < 1.0$ (top, 43 objects) and $1.0 \leq z \leq
2.3$ (bottom, 56 objects). At low (resp. high) redshifts the
probability that the distribution of polarization angles is drawn from
an uniform distribution is 0.3\% (resp. 0.2\%) with a mean value of
the polarization angles around 79$^{o}$ (resp. 8$^{o}$).}
\label{fig:mapn}
\end{figure}
\begin{figure}[!p]
\resizebox{\hsize}{!}{\includegraphics*{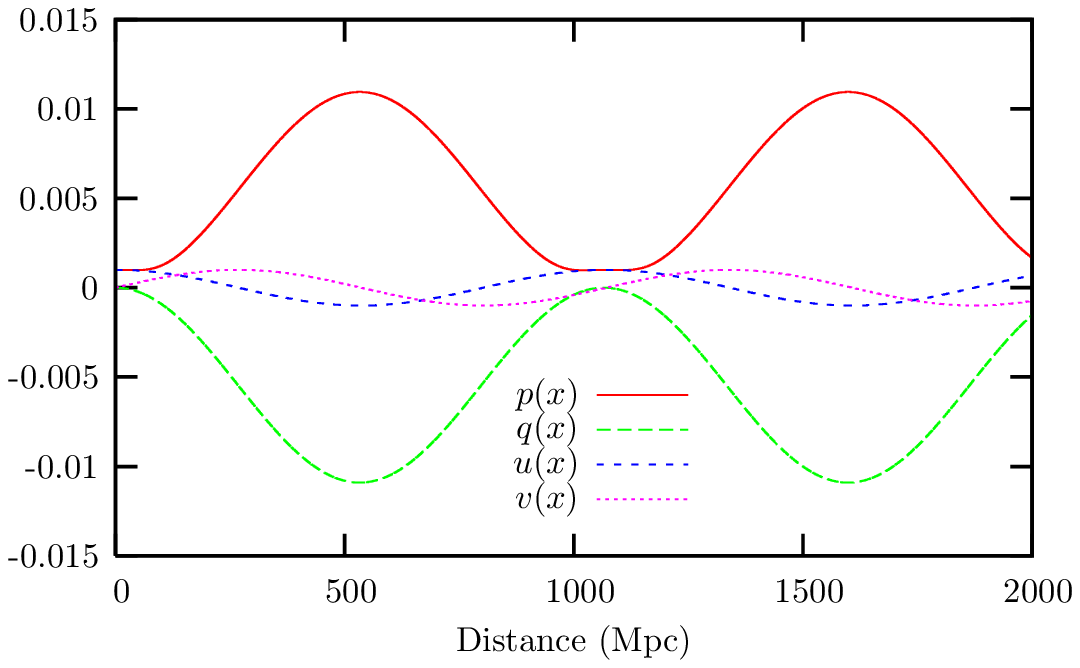}}\\
\resizebox{\hsize}{!}{\includegraphics*{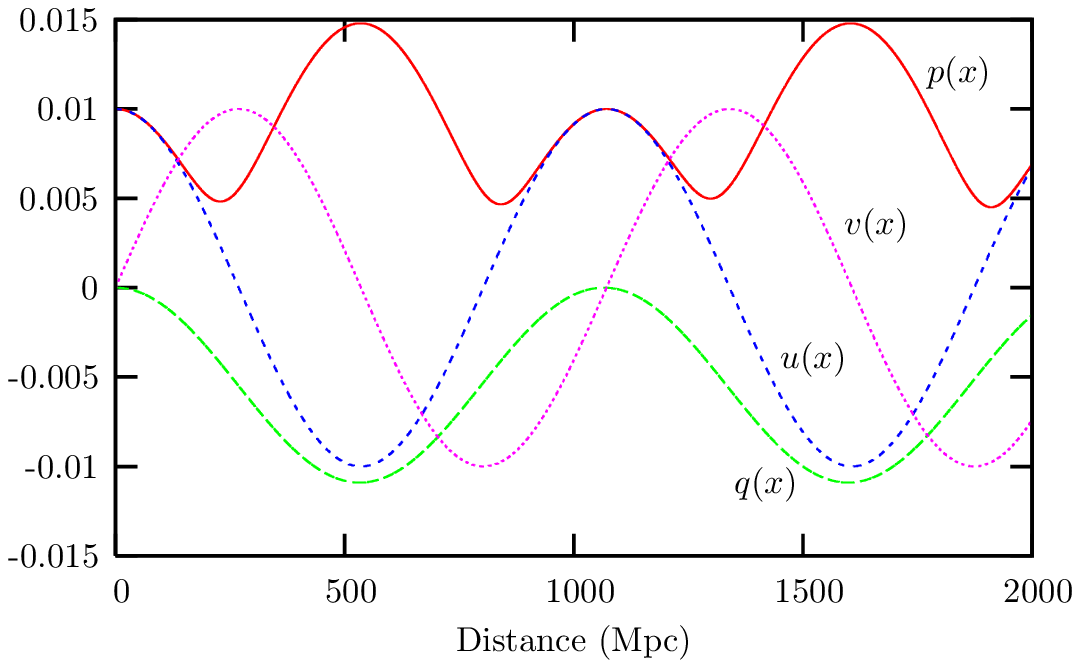}}
\caption{{\sl Top:} Evolution of the degree of linear polarization,
$p$, and of the normalized Stokes parameters $q$, $u$ and $v$ as a
function of the distance to the quasar (which corresponds to the
extent of the external magnetic field). The external magnetic field is
taken parallel to polarization vectors with $q > 0$ and $u=0$.  Among
the Stokes parameters, only $u$ has been chosen initially different
from zero, $u(0) = 0.001$. The other parameters are: the mass of the
pseudoscalar or axion-like particle $m = 3.70025\, 10^{-14}$~eV, the
photon plasma frequency $\omega_p = 3.7\, 10^{-14}$~eV, the
observation wavelength $\lambda = 500$~nm, the coupling constant
$g\simeq 7\, 10^{-12}$~GeV$^{-1}$ and the strength of the external
magnetic field $B\simeq4\,10^{-11}$~G (only the product $gB$ is
relevant).  {\sl Bottom:} Same as previous figure but for $u(0) =
0.01$.}
\label{fig:mix}
\end{figure}
The sample under study contains 355 polarized quasars up to redshifts
$z \sim 2.5$ and with optical polarization position angles ($\theta$)
reasonably well determined ($\sigma_{\theta} \leq 14^o$).  It contains
various types of bright quasars: radio-quiet and radio-loud, with or
without broad absorption lines, etc. As far as possible, blazars were
excluded due to their variable polarizations and unsecure
redshifts. To minimize contamination by interstellar polarization,
only objects with polarization degrees $p \geq 0.6\%$ and located at
high ($\geq 30^o$) galactic latitudes were considered.

Alignments of quasar polarization vectors are illustrated in
Fig.~\ref{fig:mapn} which shows maps of polarization vectors at low
($z < 1$) and high ($z > 1$) redshifts. We immediately see that the
polarization vectors are not randomly oriented as one would expect.
Circular statistics indicates that the departures from a uniform
distribution are statistically significant, and that the polarization
angles of low- and high-redshift objects cluster around significantly
different directions. The sources cover a huge region of the sky,
typically 1$\,$Gpc at $z \sim 1$.

Statistical tests based on the nearest-neighbour analysis were applied
to the full sample of 355 quasars. They indicate that the quasar
polarization vectors are not randomly distributed over the sky with a
probability in excess of 99.9\% (Paper III; Jain et
al. 2004). Coherent orientations are best detected in groups of 30-40
objects which correspond to the regions of alignments seen in
Fig.~\ref{fig:mapn}.

Possible contamination of the data by intrumental and interstellar
polarization was carefully inspected. Instrumental polarization was
measured to be very small. Furthermore, quasar polarization data
obtained at different observatories with different instruments do
agree within the uncertainties. At high galactic latitudes the
interstellar polarization is typically $\leq 0.2-0.3\%$ but can
occasionally be higher.  Although we adopt the cutoff $p \geq 0.6\%$
on the polarization degree to make sure that most of the measured
polarization is intrinsic to the quasars, the polarization of the
objects in our sample is usually lower than 2\% (Fig.~\ref{fig:mapn})
so that at least some data might be affected by interstellar
polarization. Several tests and simulations were then performed in
Paper III, as well as in Sluse et al.~(2005) and Cabanac et
al.~(2005), leading to the conclusion that it is very unlikely that
interstellar polarization can be at the origin of the observed
alignments. This is basically what we can also conclude from
Fig.~\ref{fig:mapn}: should interstellar polarization (or any local
contamination) be responsible for the observed alignments, one would
expect the same effect at all redshifts and no significant difference
between mean polarization angles at low and high redshifts.

\section{Towards an interpretation}
To summarize, current observations show that quasar polarization
vectors appear coherently oriented or aligned over huge (1 Gpc)
regions of the sky located at both low ($z \sim 0.5$) and high ($z
\sim 1.5$) redshifts and characterized by different preferred
directions of the quasar polarization. Looking in more details, there
seems to exist a regular alternance along the line of sight of regions
of randomly oriented and aligned polarization vectors, the mean
polarization angle apparently rotating with the redshift (Paper
III). Interestingly enough, the alignment effect seems to be prominent
along an axis not far from preferred directions tentatively identified
in the Cosmic Microwave Background (Ralston \& Jain 2004).

Alignments can be simulated for a group of quasars polarized at $p
\sim 2\%$ by adding a small systematic polarization ($\sim 0.5\%$) to
their intrinsic polarization vectors, assumed to be initially randomly
oriented.  This systematic polarization, which can be generated along
the line of sight, cannot be too high since existing correlations
between polarization and other morphological or spectral properties of
quasars (e.g.~Lamy \& Hutsem\'ekers 2004, Borguet et al.~2008, and
references therein) cannot be washed out.

Although various interpretations might be possible (Paper I--III; see
also Morales \& S\'aez 2007, Demia\'nski \& Doroshkevich 2007,
Campanelli et al. 2007), the dichroism and birefringence due to
photon-pseudoscalar mixing within a magnetic field can modify the
polarization of quasar light during its propagation in a way that
qualitatively reproduces the observations (Jain et al. 2002, Das et
al. 2005, Gnedin et al. 2007, Piotrovich et al. 2008, Payez et
al. 2008).

Fig.~\ref{fig:mix} illustrates the effect of photon-pseudoscalar
mixing, within an external magnetic field, on the light from a distant
quasar (cf. Payez et al. 2008 for details). A small additional
polarization which oscillates with the distance is created, such that
an alternance of regions of random and aligned polarization vectors
can be reproduced.  The wavelength dependence of the effect
(e.g. Payez et al. 2008) could explain why such alignments are not
detected at radio-wavelengths (Joshi et al. 2007).  While the adopted
magnetic field strength is lower than current upper limits for a
cosmological magnetic field, a huge coherence scale is needed. Also,
fine tuning is necessary to produce an oscillation with the observed
periodicity. While an interpretation in terms of photon-pseudoscalar
mixing seems promising, more realistic simulations are needed, such as
a study of the cumulative effect of a succession of smaller regions
with randomly oriented magnetic fields.  Interestingly, an apparently
generic effect of photon-pseudoscalar mixing is the creation of
circular polarization with $\bar{v} \simeq \bar{u}$
(Fig.~\ref{fig:mix}).

\end{document}